\documentclass[JHEP,12pt]{article}
\usepackage{jheppub}
\usepackage{amssymb,amsmath,amsthm}
\usepackage[shortlabels]{enumitem}
\usepackage{braket}
\usepackage{comment} 
\usepackage{subcaption}
\usepackage{graphicx, xcolor, varwidth}
\usepackage{color}

\theoremstyle{remark}

\DeclareMathOperator{\tr }{tr}

\author{Raphael Bousso and Geoff Penington}

\affiliation{Center for Theoretical Physics and Department of Physics,\\
University of California, Berkeley, California 94720, U.S.A. 
} 

\emailAdd{bousso@berkeley.edu}
\emailAdd{geoffp@berkeley.edu}

\title{Islands Far Outside the Horizon}

\abstract{Information located in an entanglement island in semiclassical gravity can be nonperturbatively reconstructed from distant radiation, implying a radical breakdown of effective field theory. We show that this occurs well outside of the black hole stretched horizon. We compute the island associated to large-angular momentum Hawking modes of a four-dimensional Schwarzschild black hole. These modes typically fall back into the black hole but can be extracted to infinity by relativistic strings or, more abstractly, by asymptotic boundary operators constructed using the timelike tube theorem. Remarkably, we find that their island can protrude a distance of order $\sqrt{\ell_p r_{\rm hor}}$ outside the horizon. This is parametrically larger than the Planck scale $\ell_p$ and is comparable to the Bohr radius for supermassive black holes. Therefore, in principle, a distant observer can determine experimentally whether the black hole information paradox is resolved by complementarity, or by a firewall.}

\makeatletter
\gdef\@fpheader{\mbox{}}
\makeatother

\begin{document}
\maketitle

\section{Introduction}

To preserve unitarity when a black hole forms and evaporates, the Hawking radiation must carry away all quantum information that, in semiclassical general relativity, appears to be in the interior of the black hole. This is incompatible with the algebraic structure of relativistic quantum field theory, where spacelike-separated observables commute.

Black hole complementarity \cite{Susskind:1993if, Susskind:1994vu} attempts to reconcile semiclassical physics with unitarity, by asserting that the interior and the radiation are not independent systems. This would allow the Hawking radiation to contain information about the interior, though it does not explain how the radiation evolves into such a state.

In its original incarnation, black hole complementarity required for consistency that no single observer could see two spacelike-separated incarnations of the same quantum information. This condition was shown to be violated by Almheiri \emph{et al.} (AMPS)~\cite{Almheiri:2012rt}.  Assuming (1) a unitary S-matrix for black hole formation and evaporation, (2) validity of effective field theory outside the black hole, and (3) the smooth geometry predicted by General Relativity at the black hole horizon, AMPS showed that a single computationally unbounded observer can prepare localized particles in an impossible quantum state, in a low curvature region of spacetime.

One possible response to the AMPS paradox -- and indeed the solution proposed by AMPS themselves -- is to assert that the black hole interior does not exist at all, at least after the Page time when the paradox first appears. Instead, the semiclassical black hole description is only valid outside of a ``firewall'' near the stretched horizon, where the local temperature becomes Planckian from the perspective of a static observer.

Alternatively, new forms of complementarity have been explored that rely substantially on the role of computational complexity~\cite{Harlow:2013tf, Maldacena:2013xja, Akers:2022qdl}. Sufficiently complex operations on the radiation, including those invoked in~\cite{Almheiri:2012rt} to construct the impossible quantum state, are said to change the interior state, destroying the Lorentzian causal structure of the spacetime. Absent such complex operations, the AMPS paradox cannot be operationally verified, so the horizon can be asserted to be smooth. Such approaches suffer from a number of potential ambiguities: in particular, it is unclear how to ascribe a unique experience (if one even exists) to an infalling observer in the presence of complex manipulations of the radiation that can break semiclassical causality; nor is it clear what exterior states should be considered complex enough to break causality.

The conflict between unitarity and general relativity was reshaped, without being fully resolved, by the discovery of entanglement islands~\cite{Penington:2019npb,Almheiri:2019psf}. Islands are the crucial ingredient in gravitational computations of signatures of unitarity, such as the Page curve \cite{Page:1993df, Page:1993wv}. They also allow us to determine directly from the gravitational path integral whether and when interior information is encoded in the Hawking radiation, using explicit decoding protocols such as the Petz map \cite{Cotler:2017erl, Chen:2019gbt, Penington:2019kki} or modular flow reconstruction \cite{Faulkner:2017vdd, Chen:2019iro}. Specifically, any information that  in semiclassical general relativity would pass through the island -- a particular spacetime region, usually in the black hole interior, that is bounded by a quantum extremal surface -- can be recovered from the radiation. Information that is outside the island cannot be recovered.

It is remarkable that evidence for unitarity, such as the Page curve and the encoding of the semiclassical interior in the radiation, can be derived from a calculation involving the semiclassical Hawking state, in which the horizon is smooth and information is lost. But this does not imply that the interior exists, any more than it implies that Hawking's original conclusion of information loss still holds~\cite{Bousso:2019ykv}. 

All the observables that are actually computed using entanglement islands pertain to measurements of Hawking radiation far from the black hole.
Moreover, the way the Hawking state actually arises in island calculations is rather indirect \cite{Penington:2019kki, Almheiri:2019qdq}. Typically, one first computes an observable that involves two or more copies (or ``replicas'') of the asymptotic boundary conditions leading to the black hole in question. One finds that in some cases the gravitational path integral is dominated by ``replica wormholes'' connecting the $n$ copies. Finally, to compute the quantities of primary interest, it is necessary to analytically continues the number of replicas $n \to 1$. It turns out~\cite{Lewkowycz:2013nqa, Faulkner:2013ana} that in many situations there exists a shortcut that is mathematically equivalent to the above sequence of steps: one simply finds the entanglement island (or more generally the quantum extremal region) associated to the radiation in the original spacetime, in the Hawking state. The entanglement island is the $n \to 1$ limit of the region that passes through the replica wormhole at integer $n > 1$. These fairly abstract mathematical manipulations do not necessarily tell us anything new about an infalling observer's experience. 

In particular, while entanglement islands provide the first direct evidence from the gravitational path integral for the unitarity of the exterior description, they do not tell us how to reconcile unitarity with a smooth horizon. The choices for resolving the AMPS paradox remain the same as before: either firewalls, or a new form of complementarity that restricts the complexity of external operations.

Only experiment can distinguish between these hypotheses. If we could directly observe an island, we would either confirm complementarity (assuming unitarity was also experimentally demonstrated); or we would discover a firewall.  But if the island is inside a black hole, no one can ever communicate the results of such an experiment to the outside world. (This restriction is crippling even at the level of thought-experiments in an AdS spacetime, since one cannot directly construct mathematical probes of the black hole interior using only CFT computations and the extrapolate dictionary.)

In certain situations where a black hole is entangled with infalling thermal radiation, islands may extend outside the event horizon of the black hole~\cite{Almheiri:2019yqk}. Nonetheless, as a test bed for (thought) experiments, the setting of \cite{Almheiri:2019yqk} suffers from two important drawbacks: a) the black holes considered have only two spacetime dimensions; and b) the island only extends an $O(NG)$-distance outside the horizon, where $G$ is Newton's constant and $N$ is the number of 2D matter fields. An observer who enters this island would therefore need at least Planckian acceleration to escape to infinity, and would have accessed at most an island portion of Planck width. One might try to avoid this issue by making the number of matter fields $N$ parametrically large. However, in the presence of a large number of matter fields, semiclassical locality typically breaks down at the species scale $O(NG)$~\cite{Wald:1999vt}. Thus, either way, an experiment probing the island would not be under semiclassical control. 

In this work, we consider a four-dimensional Schwarzschild black hole, which could certainly be produced by a sufficiently capable experimentalist. Happily, even in the presence of only a single scalar matter field (or gauge field), such black holes feature a parametrically large number of effective two-dimensional modes, namely Kaluza-Klein modes with large angular momentum $j$ around the transverse sphere. Furthermore, the modes with large angular momentum are almost entirely reflected off the black hole potential barrier back into the black hole; as a result, they are in thermal equilibrium with the black hole. This is exactly the situation that was found in \cite{Almheiri:2019yqk} to produce an island outside the horizon. Because of the large number of modes involved, it will turn out that the islands associated to these large-$j$ modes can extend far outside the horizon in Planck units.

Conventionally, an island $I(\mathsf{R})$ is associated to a set of QFT modes $\mathsf{R}$ (the radiation) in a weakly gravitating region far from the black hole. In fact, in formal calculations, one normally considers situations where the radiation $\mathsf{R}$ has escaped into a nongravitational quantum system (or ``bath'') that is external to the gravitating spacetime containing the black hole. This means that the algebraic structure of $\mathsf{R}$ (if not its state) is described by ordinary quantum field theory with no mysterious quantum gravitational effects.

In a black hole of radius $r_{\rm hor}$, outgoing large-$j$ modes are reflected back into the black hole at a proper distance $\rho \ll r_{\rm hor}$ outside the horizon. They almost never escape into a weakly gravitating region. An obvious question therefore is whether we are nonetheless justified in associating an island to them. Are they independent commuting QFT degrees of freedom like Hawking radiation that escapes to infinity? 

It is an unquestioned assumption in AMPS~\cite{Almheiri:2012rt} and even in standard treatments of black hole complementarity \cite{Susskind:1993if, Susskind:1994vu} that low-energy effective field (EFT) is valid at proper distances $\rho$ much larger than the Planck length $\ell_p$ outside the horizon. However, the islands we will find themselves challenge this assumption: they imply that EFT breaks down much farther from the horizon than previously assumed.

Nonetheless, the modes $\mathsf{R}$ that have islands far outside the horizon do not themselves lie in those islands. To show that the modes $\mathsf{R}$ are indeed independent degrees of freedom on the same footing as distant radiation, it suffices to exhibit
a controlled semiclassical process that extracts them all simultaneously to asymptotic infinity where the algebraic structure of the theory is clear. The backreaction from this process will place the island behind the event horizon; but this is not relevant to the question of whether the modes are independent of each other and of the distant radiation with which they are combined to obtain an island.

Indeed, by dangling a large number of strings into the near-horizon region, one can wick large-$j$ Hawking modes through the potential barrier so that they escape to asymptotic infinity~\cite{Frolov:2000kx}. This process is limited by the requirements that the backreaction from the mass of the strings not be too large and that the strings be sufficiently heavy that they are not melted by the Hawking radiation. If strings of appropriate tension are available that saturate the null energy condition, then they suffice to extract precisely the modes that we are interested in (and no more)~\cite{Brown:2012un}. More abstractly, in quantum field theory in curved spacetime, the timelike tube theorem says that modes near the black hole horizon can be extracted using operators near asymptotic infinity that are delocalised in any timeband whose causal diamond contains the near-horizon modes. (In AdS/CFT, analogous arguments are used in the context of HKLL-style bulk reconstruction to show that an operator deep in the bulk can be rewritten using bulk operators near asymptotic infinity \cite{Hamilton:2006az}.) So long as the effects of backreaction can be safely ignored, the same should be true in gravitational effective field theory. Again, the backreaction turns out to be under control for precisely the modes that we will consider below.
\begin{figure}
    \centering
\includegraphics[width=0.5\textwidth]{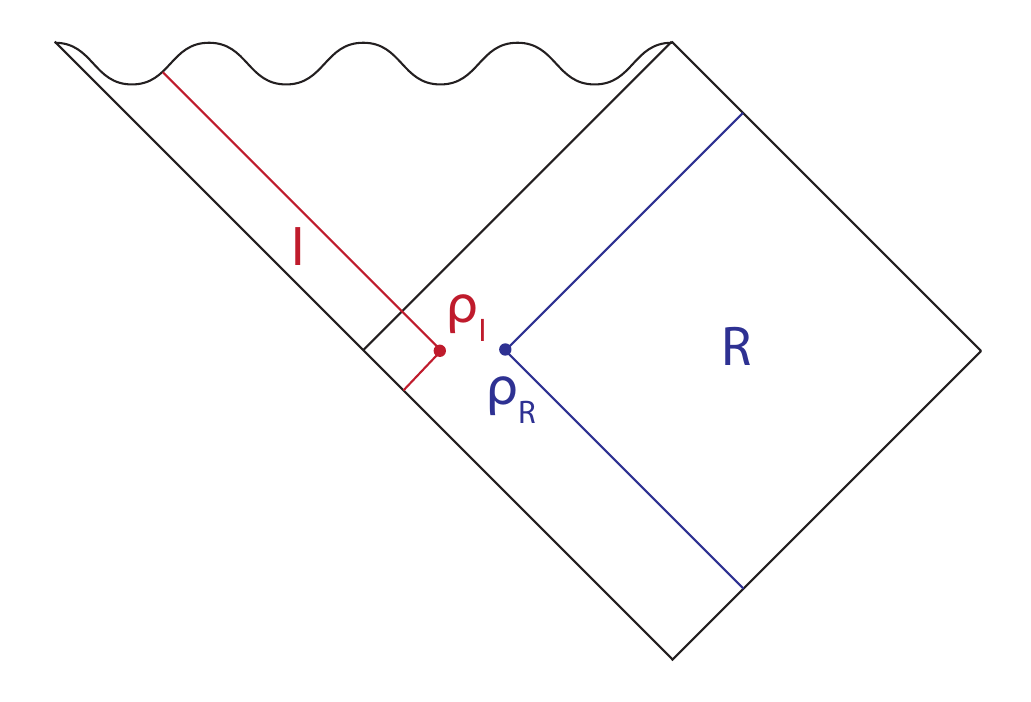}
    \caption{Effective field theory modes $\mathsf{R}$ (below at cut-off $\varepsilon_R$) that are a proper distance $\rho > \rho_R$ outside the horizon at Schwarzschild time $t = t_R$(after the Page time) have an associated entanglement island $I$. The island $I$ is bounded by a sphere a proper distance $\rho_I \ll \rho_R$ outside the horizon in the same $t=t_R$ Schwarzschild time slice.}
    \label{fig:island}
\end{figure}

Concretely, consider a black hole after the Page time, and suppose that all QFT modes $\mathsf{R}$ (below some cut-off scale) in the region $R$ defined by $t = t_R$ and $\rho > \rho_R$ are extracted to asymptotic infinity by one of the methods described above. Combined with the Hawking radiation that had propagated naturally to infinity, these modes will have an entanglement island, as shown in Figure \ref{fig:island}. Its location does not depend on the details of how the modes were extracted and can be read off directly from the state of the modes prior to extraction; the rules for doing so were first described explicitly in \cite{Dong:2020uxp}.\footnote{In recent papers \cite{Bousso:2022hlz, Bousso:2023sya}, we proposed a much more ambitious framework for associating an entanglement wedge to a region $R$ in a gravitating spacetime. In contrast to the proposal in \cite{Dong:2020uxp}, this entanglement wedge is supposed to describe the region encoded not just by the low-energy modes $\mathsf{R}$, but by a  much larger nonperturbative quantum gravity algebra associated to $R$. The calculations in the present paper have a natural interpretation in that framework. For example, they suggest that (after the Page time) the entanglement wedge of a thin annulus a distance $\rho_R$ outside the black hole only holographically encodes the region $\rho_I < \rho < \rho_R$ and not the entire black hole. However we focus on the  island associated to the low-energy modes $\mathsf{R}$ (which can be defined using only the formulas from \cite{Dong:2020uxp}) because it features all the same physics without any need to speculate about the information content of high-energy Planckian degrees of freedom.} It is therefore natural to associate this entanglement island to the Hawking modes even if they are not extracted.

We show in Section \ref{sec:location} that the island associated to QFT modes at $\rho > \rho_R$ extends a proper distance
\begin{align} \label{eq:islandlocation}
    \rho_I \sim \ell_p^2 r_\mathrm{hor}^2 \rho_R^{-3}~. 
\end{align}
outside of the black hole horizon.\footnote{After this manuscript appeared on arXiv, we became aware that the formula \eqref{eq:islandlocation} was previously obtained in a related context in \cite{Hashimoto:2020cas}. The bound $\rho_R \gtrsim \sqrt{\ell_p r_{\rm hor}} \gtrsim \rho_I$ for this solution to exist was noted in \cite{Matsuo:2020ypv}, where it was conjectured that the information in a black hole is localised on a surface saturating that bound.} For $\rho_R \ll r_\mathrm{hor}^{2/3} \ell_p^{1/3}$, this far exceeds the Planck distance. 

The smaller we choose $\rho_R$, the larger the island will be. However, the derivation of \eqref{eq:islandlocation} assumes $\rho_R \gg \rho_I$. It is therefore valid only so long as $\rho_R \gg r_{\rm hor}^{1/2} \ell_p^{1/2}$. In fact, the same condition arises independently and self-consistently as a bound on the number of modes in $\mathsf{R}$ that can be extracted to infinity using either of the methods described above; this will be shown in Section \ref{sec:mining}. 

Saturating this bound, we find that the island $I(\mathsf{R})$ can extend a proper radial distance $\rho_I^{\rm max} \sim O(r_{\rm hor}^{1/2} \ell_p^{1/2})$ outside the black hole. In an asymptotically flat spacetime, this distance can be arbitrarily large compared to the Planck scale; in principle it could be billions of lightyears. For a Schwarzschild black hole comparable in mass to known supermassive black holes, the island extends approximately a Bohr radius outside the horizon: too small for a human-sized observer to enter, but still macroscopic compared to the Planck scale.

If the black hole information problem is to be resolved via the firewall paradigm, with islands just a convenient mathematical fiction, the firewall had better extend at least as far outside the horizon as any island. Otherwise, we would have given up semiclassical physics at the horizon and yet would still have the same breakdown of the algebraic structure of quantum field theory found in black hole complementarity. We argue in Section \ref{sec-implications} that any firewall must therefore be situated at least an $O(r_{\rm hor}^{1/2} \ell_p^{1/2})$-distance outside the semiclassical location of the black hole horizon. An important consequence is that the firewall now affects boundary observables that can be directly computed from the extrapolate dictionary. Therefore the smooth bulk geometry predicted by GR cannot emerge from averaging over an ensemble of boundary theories each of which is dual to a far-out firewall.

\section{Location of the Island} \label{sec:location}
The metric of a Schwarzschild black hole in $d$ spacetime dimensions is
\begin{align} \label{eq:schwarz}
    ds^2 = - f(r) dt^2 + f(r)^{-1} dr^2 + r^2 d\Omega^2~,
\end{align}
where $f(r) = 1 - r_{\rm hor}^{d-3}/r^{d-3}$. This metric accurately describes a black hole formed from collapse at time $t_{\rm coll} \ll 0$ for $r - r_{\rm hor} \gg r_{\rm hor} \exp(- 4\pi (t  - t_{\rm coll})/\beta)$, where $\beta = 4 \pi/f'(r_{\rm hor})$  is the inverse temperature of the black hole. To have a nontrivial island, the black hole needs to be after the Page time, i.e., $t_{\rm coll} = - O(S_{BH}\beta)$, where $S_{BH} =A_{\rm hor}/4G$ is the Bekenstein-Hawking entropy. So \eqref{eq:schwarz} is valid so long as $(r - r_{\rm hor})$ is not exponentially small. In the near horizon region, $r-r_{\rm hor}\ll r_{\rm hor}$, we have
\begin{align} \label{eq:nearhormetric}
    ds^2 = - \frac{(d-3)^2}{4\, r_{\rm hor}^2} \rho^2 dt^2 + d\rho^2 + \left(r_{\rm hor} + \frac{d-3}{4\, r_{\rm hor}} \rho^2\right)^2 d\Omega^2~,
\end{align}
where $\rho = 2\sqrt{r_{\rm hor}(r - r_{\rm hor})/(d-3)}$ is the proper distance from the horizon on a hypersurface of constant $t$.  

The island rule \cite{Penington:2019npb, Almheiri:2019psf, Almheiri:2019hni},\footnote{The island rule is really a special case of a more general rule for holographic entropies called the quantum extremal surface (QES) prescription \cite{Ryu:2006bv,Hubeny:2007xt, Wall:2012uf, Engelhardt:2014gca, Hayden:2018khn}.} applied to low-energy quantum field theory modes $\mathsf{R}$ (below some cut-off energy scale $1/\varepsilon_R$) in a gravitating spacetime \cite{Dong:2020uxp}, says that the entropy of those modes is given by
\begin{align}
S(\mathsf{R}) = \underset{I}{\rm min ext}\,\, S_\mathrm{gen} (\mathsf{R} \cup I) = \underset{I}{\rm min ext}\,\, \left[\frac{A(\eth I)}{4G} +  S_\mathrm{bulk} (\mathsf{R} \cup I)\right]~.
\end{align}
Here the generalised entropy $S_\mathrm{gen} (\mathsf{R} \cup I)$ decomposes into a piece proportional the area $A(\eth I)$ of the edge of the island and  the von Neumann entropy $S_{\rm bulk}(\mathsf{R} \cup I)$ of QFT modes in $R \cup I$ computed as if we were working in pure QFT with no gravity; the terminology $S_{\rm bulk}$ for the latter quantity comes from AdS/CFT. Importantly, there is an explicit cut-off $\varepsilon_R$ on the modes in $R$, but not on the modes in $I$: this leads a divergence in $S_{\rm bulk}$ that is renormalised by the area term. The region $I$, which should be spacelike separated from $R$, is found by looking for regions $I$ such that the generalised entropy $S_\mathrm{gen} (\mathsf{R} \cup I)$ is constant at linear order with respect to small perturbations of $\eth I$; if multiple such regions exist the one with smallest generalised entropy is chosen.

Rindler modes with angular momentum $j$ are reflected back into the black hole around $\rho \sim r_{\rm hor}/j$ where their wavelength in local Minkowski coordinates is of order $\rho$. We therefore define the region $R$ defined as the domain of dependence of the set of points with $t = 0$ and $\rho > \rho_R$ for some $\rho_R \ll r_{\rm hor}$ and choose a cut-off wavelength for modes to be included in $\mathsf{R}$ to be $\varepsilon_R = \varepsilon \rho_R$ with some fixed $\varepsilon\ll 1$. After the Page time, the modes $\mathsf{R}$ should encode an entanglement island bounded by a sphere at fixed radius and Schwarzschild time that is (relatively) near the horizon. 

To a very good approximation, we can compute the location of the edge of the island by pretending that the near-horizon region is in the Hartle-Hawking state. This approximation is correct up to exponentially small errors for all near-horizon modes except for the small number of Hawking modes with $O(1)$ angular momentum that can tunnel through the potential barrier and escape to infinity. We shall later verify that the effect of those modes on the location of the island is small.

 An advantage of working with the Hartle-Hawking state is that it has a time-reflection symmetry. As a result, the edge $\eth I$ of the island must be a constant-radius sphere in the Schwarzschild $t = 0$ slice.
To find the location of the island, we therefore only need to compute the generalized entropy 
\begin{align}
    S_{\rm gen}(\mathsf{R} \cup I) = \frac{A(\eth I)}{4G} + S_{\rm bulk}(\mathsf{R} \cup I)
\end{align} 
of candidate islands of this form as a function of the radial location $\rho_I$ of their edge. The location $\rho_I$ of the edge of the island can then be found by solving for
\begin{align} \label{eq:extremal}
\frac{d}{d\rho_I}  S_{\rm gen}(\mathsf{R} \cup I) = 0~.
\end{align}
Together with the symmetries of the Hartle-Hawking state, \eqref{eq:extremal} is sufficient to ensure that the island is extremal.

We first compute the area gradient as a function of $\rho_I$. In the near horizon region $\rho_I \ll r_{\mathrm{hor}}$, it follows immediately from \eqref{eq:nearhormetric} that
\begin{align}
    \frac{d A(\eth I)}{d \rho_I} = \frac{(d-2)(d-3)\pi^{(d-1)/2} }{\Gamma(\frac{d-1}{2})} \,r_{\mathrm{hor}}^{d-4} \rho_I~.
\end{align}
To compute the entropy gradient, we note that, since the global state is pure, we have
\begin{align}
S_{\rm bulk}(\mathsf{R} \cup I) = S_{\rm bulk}(B)~,
\end{align}
where $B$ is the thin shell defined by $\rho_I < \rho < \rho_R$. Since the width of the shell $B$ is parametrically smaller than than the black hole radius $r_{\rm hor}$, $S_{\rm bulk}(B)$ can be approximated using standard formulas for a thin slab of width $(\rho_R - \rho_I)$ and cross-sectional area $A_{\rm hor} = r_{\rm hor}^{d-2}$. The entropy of such a slab is given by
\begin{align}\label{eq:sbulk}
  S_{\rm bulk}(\mathsf{R} \cup I) = S_{\rm bulk}(B) = \frac{A_{\rm hor}}{\varepsilon_I^{d-2}} + \frac{A_{\rm hor}}{\varepsilon_R^{d-2}} - \kappa \frac{A_{\rm hor}}{(\rho_R - \rho_I)^{d-2}}~.
\end{align}
The divergent first term in \eqref{eq:sbulk} comes from UV-modes near the edge of $I$. It is absorbed into the area term in $S_{\rm bulk}(\mathsf{R} \cup I)$ via the renormalization of Newton's constant. The second term comes from UV-modes near the edge of $R$ and depends on the physical cut-off scale $\varepsilon_R$ in the definition of the modes $\mathsf{R}$. The finite final term is the only part that depends on $\rho_I$ and hence affects the location of the island. The prefactor $\kappa$ is an $O(1)$ universal constant that depends on the details of the bulk fields present and the spacetime dimension. For a single free massless scalar field in four dimensions, we have $\kappa \approx 0.0049$. For holographic conformal field theories, we have
\begin{align}
    \kappa = \frac{2^{d}\pi^{d/2} \Gamma\left(\frac{d+1}{2d}\right)^d}{4 G_{AdS} (d-1)  \Gamma\left(\frac{1}{2d}\right)^d}~,
\end{align}
where $G_{AdS}$ is Newton's constant in AdS units in the dual bulk theory.

The physically relevant solutions to \eqref{eq:extremal} have $\rho_I \ll \rho_R$.\footnote{Since \eqref{eq:sbulk} diverges to negative infinity as $\rho_I \to \rho_R$ while the area $A(\eth I)$ stays finite, there is also formally a solution of \eqref{eq:extremal} for $\rho_R \gg O(G^{1/d}r_{\rm hor}^{2/d})$ that has $\rho_R - \rho_I \ll \rho_R$. This solution scales as $\rho_R - \rho_I  = O\left(r_{\rm hor}^{2/(d-1)} G^{1/(d-1)}/\rho_R^{1/(d-1)}\right)$.
However, as we shall see in Section \ref{sec:mining}, one cannot extract the near-horizon modes (and hence trust that the island makes sense) unless the cut-off $\varepsilon_R \gg G^{1/d} r_\mathrm{hor}^{2/d}$. The formal solution with $\rho_R - \rho_I \ll \rho_R$ therefore has $\varepsilon_R \gg \rho_R - \rho_I$ and hence the formula \eqref{eq:sbulk} is not valid. 
}
In this limit, \eqref{eq:extremal} becomes
\begin{align} \label{eq:solveisland}
    \frac{d}{d\rho_I}  S_{\rm gen}(\mathsf{R} \cup I) = \frac{(d-2)(d-3)\pi^{(d-1)/2} }{4 G \Gamma(\frac{d-1}{2})} \,r_{\mathrm{hor}}^{d-4} \rho_I - \kappa (d-2) \frac{A_{\rm hor}}{\rho_R^{d-1}} = 0~,
\end{align}
and hence the island is located at
\begin{align} \label{eq:islandlocationprecise}
    \rho_I = \frac{8 \kappa G r_\mathrm{hor}^2}{(d-3) \rho_R^{d-1}}~. 
\end{align}

The distance $\rho_I$ of the island outside the horizon is maximized by making the radius $\rho_R$ as small as possible subject to the constraint $\rho_R \gg \rho_I$ (i.e., by making $\rho_R$ and $\rho_I$ scale identically in the limit $r_{\rm hor}/\ell_p \to \infty$. In such a limit, we have
\begin{align} \label{eq:maxislanddistance}
    \rho_R^{\rm min} \sim \rho_I^{\max} \sim O(G^{1/d}r_{\rm hor}^{2/d}) \sim O(\ell_p^{(d-2)/d} r_{\rm hor}^{2/d})~.
\end{align}
We will see in Section \ref{sec:mining} that the same bound on $\rho_R$ can be found by demanding that all the modes in $\rho_R$ can be extracted to infinity. In four dimensions, this gives the scaling $\rho_I \sim \sqrt{\ell_p r_{\rm hor}}$ stated in the introduction.

We now analyze the effects of the Hawking modes with $j= O(1)$ angular momentum and show that it is indeed small. These modes preserve the rotational symmetry of the Schwarzschild solution but break the time-reversal symmetry. It is therefore necessary to verify that the perturbation $\delta \rho_I, \delta t_I$ of both the radial coordinate $\rho_I$ and the Schwarzschild time $t_I$ of the edge of the island from those modes is small in order to trust the formula \eqref{eq:islandlocationprecise}. We have
\begin{align}
    \partial_{\rho_I} S_{j = O(1)}(R \cup I) \sim \frac{r_{\rm hor}}{\rho_I}\partial_{t_I} S_{j = O(1)}(R \cup I) \sim O\left(\rho_R^{-1}\right),
\end{align}
where $S_{j = O(1)}(R \cup I)$ is the contribution to the bulk entropy from the small-$j$ modes. On the other hand
\begin{align}
\nabla^2_{\rho_I} S_{\rm gen}(\mathsf{R} \cup I) \sim \frac{r_{\rm hor}^2}{\rho_I^2}\nabla^2_{t_I} S_{\rm gen}(\mathsf{R} \cup I) \sim O\left(\frac{r_{\rm hor}^{d-4}}{G}\right)~,
\end{align}
since the dominant contribution comes from classical area term $A(\eth I)/4G$. (Here $\nabla$ means the covariant derivative.) Meanwhile $\nabla_{\rho_I} \nabla_{t_I} S_{\rm gen}(\mathsf{R} \cup I)$ vanishes at this order.
It follows that the perturbation in the location of the island scales as
\begin{align}
    \delta \rho_I \sim \frac{\rho_I}{r_{\rm hor}} \delta t_I \sim O\left(\frac{G}{\rho_R r_{\rm hor}^{d-4}}\right).
\end{align}
We see that $\delta \rho_I \ll \rho_I$ and $\delta t_I \sim \rho_R^{d-2}/r_{\rm hor}^{d-3} \ll \rho_R$. Both are small compared to the scales we are interested in.
\begin{figure}
    \centering
\includegraphics[width=0.5\textwidth]{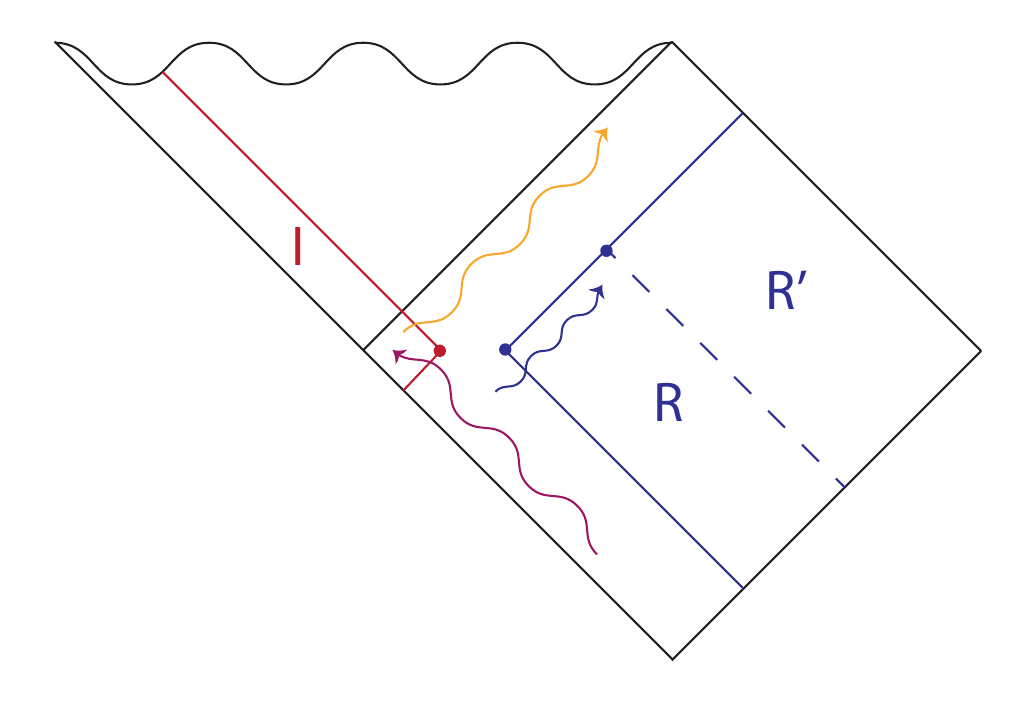}
    \caption{An outgoing Hawking mode (blue) enters the region $R$ that extends near to the horizon at an earlier Schwarzschild time than a region $R'$ farther from the horizon. Combined with a correction from the presence of a large number of (large-$j$) Hawking modes in $\mathsf{R}$, this effect reduces the time delay required for an ingoing mode (red) to become encoded in $\mathsf{R}$ predicted by the Hayden-Preskill decoding criterion. A time-reversed version of the same argument fixes the outgoing modes (orange) that should be encoded in $\mathsf{R}$. Both ingoing and outgoing modes must pass through the island associated to $\mathsf{R}$ if and only if they are encoded in $\mathsf{R}$. This fixes the location of the island up to $O(1)$ factors.}
    \label{fig:haydenpreskill}
\end{figure}

Even without knowledge of the island rule, the approximate location of the island for Hawking radiation far from the black hole can be derived by assuming only unitarity and the Hayden-Preskill decoding criterion \cite{Hayden:2007cs} for the time at which ingoing modes become in encoded in the radiation \cite{Penington:2019npb}. The approximate scaling of \eqref{eq:islandlocationprecise} also follows from the same two assumptions.\footnote{The Hayden-Preskill criterion follows from the fast scrambling dynamics of the black hole. In the context of AdS/CFT, this can be seen from simple bulk computations such as out-of-time-order correlation (OTOC) functions. In other spacetimes, gravitational OTOCs behave the same way, but the interpretation in terms of fast scrambling unitary dynamics of holographic degrees of freedom is less clear because of the lack of an explicit holographic dual.} The Hayden-Preskill decoding criterion says that we should expect information falling into the black hole to become encoded in the radiation after waiting for a Schwarzschild time $ \Delta t = (\beta/2\pi) \log S_{BH}$. However in our context this timescale is reduced by two effects. Firstly, radiation reaches a small distance $\rho$ outside the horizon a Schwarzschild time $\Delta t = (\beta/2\pi) \log (r_{\rm hor}/\rho) $ before it escapes the black hole entirely (i.e., reaches $\rho \sim r_{\rm hor}$). So the smaller we make $\rho_R$ the sooner the same information enters $R$ just because of the relativistic dynamics of the outgoing Hawking radiation. Secondly, collecting a large number of modes $N$ emitted at the same time reduces the required delay by $(\beta/2\pi) \log N$ because the information does not need to have been fully scrambled into the microscopic black hole degrees of freedom for it to influence the state of $O(N)$ modes \cite{Penington:2019npb}. In our case, we have $N \sim j_{\rm max}^{d-2} \sim (r_{\rm hor}/\rho)^{d-2}$. 

The net result of combining these two adjustments to Hayden-Preskill is that any information that fell into the black hole at Schwarzschild times 
\begin{equation}
    t < - \frac{\beta}{2\pi} \log \frac{\rho^{d-1}}{ r_{\rm hor} G }
\end{equation}
should be encoded in $\mathsf{R}$ at $t=0$. But, by a time-reversed version of the same argument, we should also expect that any information that will escape the black hole at $t > (\beta/2\pi) \log (\rho^{d-1}/ r_{\rm hor} G )$ is encoded in $\mathsf{R}$ at $t=0$. As illustrated in Figure \ref{fig:haydenpreskill}, combining these two results is sufficient to conclude that any information that semiclassically would be located at $\rho \lesssim G \rho_{\rm hor}^2/\rho_R^{d-1} \sim \rho_I$ should be encoded in the radiation  $\mathsf{R}$. Up to calculating the correct $O(1)$ prefactors, this is exactly the result we found in \eqref{eq:islandlocationprecise}.

We close by noting that the generalized entropy of the island gives matches the entropy expect for $\mathsf{R}$ in a unitary Page curve. Let $S_{\rm gen}^{\infty}\approx A_{\rm hor}/4G$ be the generalized entropy of the distant radiation and its island, i.e., the entropy of the distant Hawking radiation in the fundamental description after the Page time. The fundamental entropy should decrease when modes in the black hole's thermal atmosphere are adjoined to the distant radiation system, approximately by the number of such modes. Thus, unitarity requires that 
\begin{equation}
    S_{\rm gen}(\mathsf{R}\cup I) = S_{\rm gen}^\infty - O\left(\frac{r_{\rm hor}^{d-2}}{\rho_R^{d-2}}\right)~.
\end{equation}
Indeed, we find from \eqref{eq:sbulk} and \eqref{eq:islandlocationprecise} that 
\begin{equation} \label{eq:sgenchange}
    S_{\rm gen}(\mathsf{R} \cup I) =  - \kappa  \frac{A_{\rm hor}}{\rho_R^{d-2}} + O(r_{\mathrm{hor}}^{d-4} \rho_I^2) + \dots~,
\end{equation}
where we have dropped terms that are independent of $\rho_R$. Note that only the subleading $O(r_{\mathrm{hor}}^{d-4} \rho_I^2)$ piece depends depends directly on the location of the edge of the island: the anticipated leading-order decrease in generalized  entropy from including near-horizon modes in $\mathsf{R}$ just comes from the direct effect of the change in $\rho_R$ on the bulk entropy $S_{\rm bulk}(\mathsf{R} \cup I)$.

\section{Mining and Backreaction} \label{sec:mining}

If we want to be confident that the island associated to large-$j$ Rindler modes in the near-horizon region $\rho > \rho_R$ has the same physical interpretation as the island associated to Hawking radiation that escapes to infinity, it is important that all the Rindler modes that make it to $\rho > \rho_R$ can, in principle, be extracted to infinity.  

There is a long history of work examining the extent to which this is true. The original proposal for black hole mining suggested in fact that it should  be possible to extract all modes that reach a proper distance $\rho \gg \ell_p$ from the horizon \cite{Unruh:1982ic, unruh1983mine}. This would be inconsistent with the results in Section \ref{sec:location}, because one could simultaneously extract modes in the island, and all the radiation they were supposed to be encoded in, to infinity. After doing so, one would have as much time as needed  to decode the modes $\mathsf{R}$ and extract some quantum information that escaped from the island to infinity. Since all the modes involved would be at infinity, it would be completely safe to ignore the effects of gravity, and there would be no way for any form of black hole complementarity to save us from this quantum cloning paradox.\footnote{In AdS/CFT, for example, the modes could all be extracted into a nongravitational bath before the decoding process began. As a result it would be completely impossible by the axioms of QFT for the decoding to affect the state of the island modes as would be required to avoid a paradox. This argument assumes that the method used to extract modes in the island is independent of the state of the radiation $\mathsf{R}$, otherwise the information encoded in the radiation could be changed by the extraction process. In fact, in AdS/CFT all sub-Planckian QFT modes in the black hole exterior can be extracted into a bath using the techniques of entanglement wedge reconstruction, but the extraction of the island necessarily depends on the state of the radiation for the reasons described in \cite{Hayden:2018khn, Akers:2019wxj, Akers:2021fut}. }

However the analysis in \cite{Unruh:1982ic, unruh1983mine} ignored the mass, and hence the backreaction, of the machinery required to mine the black hole and, in particular, of the strings required to raise and lower that machinery into the near horizon region \cite{Brown:2012un}.

The null energy condition (NEC) says that the tension $T$ of a string is bounded by $T \leq \sigma$, where $\sigma$ is the mass of the string per unit length. This bound is saturated, for example, by (classical) fundamental strings in string theory. A NEC-saturating string of constant mass density stretching from the horizon to infinity can exactly support its own weight; the decrease in required tension from gravitational redshift as one ascends the string is exactly cancelled by the increased mass of string below that needs to be supported.\footnote{Strings made out of more traditional materials such as carbon nanotubes, steel or actual string instead have a maximum tension $T = w\sigma$ for some material dependent constant $w \ll 1$ and are almost completely useless by comparison. In the near-horizon region, the mass of such a string needs to grow exponentially just in order to support its own weight. As a result, it is impossible even to probe $\rho \sim r_{\rm hor}$ unless we have at least $w\sim O(1)$.} If the string stops at a distance $\rho$ above the horizon it can therefore support at most the $O(\sigma \rho)$ weight of the ``missing'' string below.
\begin{figure}
    \centering
\includegraphics[width=0.5\textwidth]{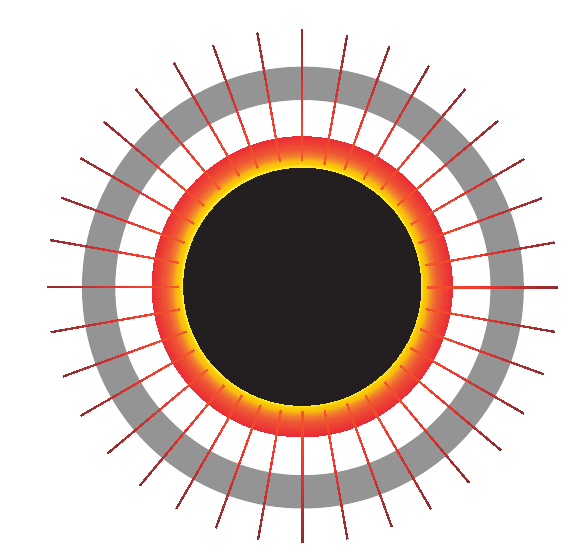}
    \caption{NEC-saturating strings threading the near-horizon region of a black hole will wick large-$j$ Hawking modes through the potential barrier (grey) and out to infinity.}
    \label{fig:mining}
\end{figure}

In fact, if you have NEC-saturating strings, you can efficiently mine a black hole without any further machinery. Simply the presence of strings stretching from the near-horizon region to infinity will ``wick'' large-$j$ modes through the potential barrier to infinity, as shown in Figure \ref{fig:mining} \cite{Frolov:2000kx}.\footnote{More complicated constructions such raising and lowering boxes filled with radiation cannot do better than this simple approach \cite{Brown:2012un}.} The main limitation on black-hole mining in this way is that each string can only wick up at most roughly one Hawking quantum per thermal time $\beta$. And to avoid large backreaction, the number of strings $N_s$ is bounded by $O(\sigma^{-1} r_{\rm hor}^{d-4} G^{-1} )$. As a result, for optimal extraction the string mass $\sigma$ needs to be as light as possible. However, if the strings are too light, the blueshifted Hawking radiation will inevitably cause them to melt when the local Hawking temperature reaches $T_R \sim \rho_R^{-1} \sim \sqrt{\sigma}$. Imposing this additional constraint means that the number of Hawking modes that can be extracted is \cite{Brown:2012un} 
\begin{align}
N  \sim \frac{r_{\rm hor}^{d-2}}{\rho_R^{d-2}} \lesssim N_s\ll  \frac{r_{\rm hor}^{(d-2)^2/d}}{G^{(d-2)/d}}~.
\end{align} 
This is exactly the bound on $\rho_R$ we found in \eqref{eq:maxislanddistance} from the assumption that $\rho_R \gg \rho_I$. Note that the problem here comes when you try to extract \emph{all} Hawking modes that reach $\rho_R$. If we only want to extract a single mode (potentially with very large $j$), we can use heavier strings and hence extract modes from deeper without running into issues with strings melting \cite{Almheiri:2012rt}.

The same constraint also appears if why try to argue more abstractly -- based on general principles of quantum field theory -- that the modes $\mathsf{R}$ can be extracted to infinity. To understand this argument, it is helpful to first consider the case where we set $\rho_R = \epsilon r_{\rm hor}$ for some small but fixed $\epsilon \ll 1$ as we take the limit $G \to 0$ (and only afterwards take $\epsilon \to 0$). 

In this limit, the number $O(\epsilon^{-d+2})$ of near-horizon Hawking modes $\mathsf{R}$ is large, and so the approximations made in order to derive \eqref{eq:islandlocationprecise} are all valid. However, we have $\rho_I \sim G r_{\rm hor}^{-d+3} \epsilon^{-d+1}$, which is much smaller than the Planck length for $d>3$. So the island extends outside, but not far outside, the horizon. 

On the other hand, the limit has the nice property that the Hawking modes are described by pure quantum field theory in a fixed curved background. Given any cut-off $\varepsilon_R$ (again small but fixed in units of $r_{\rm hor}$) we can define a continuum QFT isometric operator $V_{\rm ext}$ that extracts all QFT modes $\mathsf{R}$ into some external quantum reference system $X$. The timelike tube theorem \cite{borchers1961vollstandigkeit, araki1963generalization, Strohmaier:2023opz} in algebraic QFT  says that $V_{\rm ext}$ can be arbitrarily well approximated by a (nonlocal) operator $\tilde V_{\rm ext}$ at asymptotic infinity on any timeband that includes $R$ in its causal diamond. So, for example in asymptotically-AdS spacetimes, we obtain a bulk QFT operator acting at the asymptotic timelike boundary, while in asymptotically flat spacetimes we obtain an operator acting at past and future null infinity.

If $\rho_R/ r_{\rm hor} \to 0$ at the same time as $G \to 0$, as needed to find islands far outside the horizon, we have to be slightly more careful. Formally, even at finite $G$, we can just ignore backreaction and define an operator $\tilde V_{\rm ext}$ at asymptotic infinity that would extract the modes $\mathsf{R}$ if there was no gravity. The operator $\tilde V_{\rm ext}$ then makes sense even at finite $G$ because asymptotic boundary operators are gauge invariant in quantum gravity. However we are not guaranteed that $\tilde V_{\rm ext}$ acts the same way as $V_{\rm ext}$ on our state of interest, once the effects of backreaction are included.

A necessary condition for $\tilde V_{\rm ext}$ to act correctly is that the state where the modes $\mathsf{R}$ have been extracted can satisfy Einstein's equations with backreaction that is small compared to the relevant scales of interest. For example, the shift in the location of the event horizon should be small compared to $\rho_R$.\footnote{It is not quite obvious that this is also a sufficient condition for $\tilde V_{\rm ext}$ to act correctly. For example, if $U(0)$ is a local boundary unitary at $t=0$ and $V(t_{\rm scr})$ is a local boundary unitary a scrambling time in the future then we have $V(t_{\rm scr})^\dagger U(0) V(t_{\rm scr}) \ket{\psi} \approx U(0) \ket{\psi}$ in quantum field theory, but not in semiclassical gravity, even though the backreaction in the state $U(0) \ket{\psi}$ seems irrelevant, because there would be significant backreaction in the intermediate state $U(0) V(t_{\rm scr}) \ket{\psi}$. Since $\tilde V_{\rm ext}$ only involves operators at infinity separated by much less than a scrambling time, we do not expect that anything like that would happen here; in free field theories, we show explicitly below that it does not.} By dimensional analysis, extracting the modes $\mathsf{R}$ creates a positive energy shock density of order $\varepsilon_R^{-d}$ in a local frame
at $\rho_R$ with proper width $O(\epsilon_R)$.
Hence the Ricci tensor of the backreaction is $R_{\mu \nu} \sim G \varepsilon_R^{-d}$. This is small compared to the natural length scale of the original Schwarzschild solution so long as 
\begin{align} \label{eq:cutoffbound}
\varepsilon_R \gg G^{1/d}r_{\rm hor}^{2/d}~.    
\end{align} 
Since we must preserve $\epsilon_R\ll \rho_R$ in order to extract the relevant modes, this condition reduces to the previous constraint on $\rho_R$. In particular, one can show using Raychaudhuri's equation that the perturbative backreaction $\delta \rho_{\rm hor}$ of this energy momentum on the location of the horizon scales as
\begin{align}
    \delta\rho_{\rm hor} \sim \frac{G r_{\rm hor}^2}{\varepsilon_R^{d-1}}~.
\end{align}
As a result, the backreaction of extracting the radiation is always sufficient to hide the island behind the horizon. This avoids any paradoxes where both the island and the radiation encoding it are able to escape to infinity. However the backreaction is small compared to $\rho_R$ so long as $\varepsilon_R \sim \rho_R$ and $\rho_I \ll \rho_R$.

Indeed, as was pointed out in \cite{Almheiri:2019yqk}, the fact that the island becomes hidden behind the horizon whenever the radiation is extracted to infinity follows directly from the quantum focusing conjecture \cite{Bousso:2015mna}. A lightlike deformation of a quantum extremal surface cannot increase its generalized entropy. If the region $R$ extends close to the horizon, this fact does not tell us anything terribly useful about the location of the QES relative to the horizon, because we would expect $S_{\rm gen}(R \cup I)$ to diverge to negative infinity as $I$ becomes lightlike separated from $R$. However, if the information in $\mathsf{R}$ has been extracted to future null infinity, then the lightsheet from the QES can be extended until either it hits the singularity (if inside the horizon) or reaches asymptotic infinity (if outside the horizon). In the latter case,\footnote{A careful analysis shows that this is true even if the QES is only partially outside the horizon.} the area term contribution would cause the generalized entropy of the lightsheet to become large, in contradiction with the quantum focussing conjecture.

In free field theories, we can describe the operator $\tilde V_{\rm ext}$ somewhat more explicitly. We focus on a single large-$j$ near-horizon wavepacket with creation and annihilation operators $a^\dagger_{\rm near}$ and $a_{\rm near}$ respectively.\footnote{Since large-$j$ modes are almost perfectly reflected by the potential barrier, it does not matter here whether $a_{\rm near}^\dagger$ and $a_{\rm near}$ describe an outgoing Hawking mode or its reflection falling back into the black hole; the two are the same up to exponentially small corrections.} In a free theory, the operator $ V_{\rm ext}$ that extracts all modes $\mathsf{R}$ can be written as a product over extraction operators $V_{\rm ext}^{(1)}$ that each extract a single wavepacket.

Let $b\dagger$, $b$ be creation and annihilation operators for an external quantum system $X$, with vacuum state $\ket{0}$.\footnote{This system should be a simple harmonic oscillator if $a_{\rm near}$ is bosonic, or a fermion if $a_{\rm near}$ is fermionic} We can define $V_{\rm ext}^{(1)}$ by 
\begin{align}
    V_{\rm ext}^{(1)} \ket{\Psi} = \exp\left( \frac{\pi}{2}\left[b_X^\dagger a_{\rm near} - a_{\rm near}^\dagger b_X \right] \right) \ket{\Psi}\ket{0}_X.
\end{align}
Here $\ket{\Psi}$ is the full state of the gravitational spacetime, which includes as a subsystem the near-horizon mode acted on by $a_{\rm near}, a_{\rm near}^\dagger$. It is easy to check that $V_{\rm ext}^{(1)} \ket{\Psi} = \ket{\Psi} \ket{0}$ if $a_{\rm near} \ket{\psi} = 0$ (i.e., the near-horizon mode is in its ground state) and that
\begin{align}
    \left[b^\dagger_X a_{\rm near} - a_{\rm near}^\dagger b_X , \,a_{\rm near}^\dagger\right] = b^\dagger_X ~,~~~~~~~\left[b^\dagger_X a_{\rm near} - a_{\rm near}^\dagger b_X , \,b^\dagger_X\right] = -a_{\rm near}^\dagger~.
\end{align}
Hence the effect of $\exp\left( \frac{\pi}{2}\left[b^\dagger_X a_{\rm near} - a_{\rm near}^\dagger b_X \right] \right)$ is to ``rotate'' excitations of $a_{\rm near}^\dagger$ into $b^\dagger_X$.

It remains to rewrite $V_{\rm ext}^1$ in terms of operators near asymptotic infinity (as would be required to define the operator $\tilde V_{\rm ext}$ in gravitational EFT). The operator $a_{\rm near}$ is linearly related to the annihilation operators $a_{\rm in}$ and $a_{\rm out}$ for the corresponding ingoing and outgoing modes near asymptotic infinity by the transfer matrix for the one-dimensional quantum mechanics scattering problem. In the limit where the potential barrier is large (i.e., large $j$), we have
\begin{align}
    a_{\rm near} \approx \frac{1}{\sqrt{T}}\left[a_{\rm in} - a_{\rm out}\right]~,
\end{align}
where $T$ is the probability of transmission through the potential barrier and we have absorbed any phases into the definitions of $a_{\rm in}$ and $a_{\rm out}$. To extract the near-horizon mode, we therefore have to implement the unitary
\begin{align}
    U_{\rm ext} = \exp\left( \frac{\pi}{2\sqrt{T}}\left[b_X^\dagger a_{\rm in} - b_X^\dagger a_{\rm out}- a_{\rm in}^\dagger b_X + a_{\rm out}^\dagger b_X \right] \right)~.
\end{align}
Using standard techniques for Hamiltonian simulation (e.g.\ Trotterisation), this can be implemented using a product of an $O(T^{-1})$ (for any $\delta > 0$) number of quasilocal unitaries at asymptotic infinity, each acting only on either the ingoing or outgoing asymptotic mode. At any intermediate stage of the process, the occupation number of each mode is $O(1)$ and so no backreaction issues arise. We can therefore define a corresponding operator $\tilde V_{\rm ext}^{(1)}$ in the gravitational EFT without issue.

The transmission probability $T$ scales as $\exp[-O(j^2)] \sim \exp[-O(r_{\rm hor}^2/\rho_R^2)]$. The complexity of extracting each near-horizon mode is therefore very large, but still small compared to the $\exp[O(r_{\rm hor}^{d-2}/G)]$ complexity that is expected to (potentially) break down effective field theory and violate semiclassical causality in modern versions of black hole complementarity \cite{Harlow:2013tf, Maldacena:2013xja,Akers:2022qdl}.

It is worth noting briefly that the existence of the operator $\tilde V_{\rm ext}$ is very closely related to the generalized entropy $S_{\rm gen} (\mathsf{R} \cup I)$ having a sensible interpretation as the entropy $S(\mathsf{R})$ of the modes $\mathsf{R}$. Suppose we have $n$ copies of our gravitational state $\ket{\Psi}$. If we can define the operator $\tilde V_{\rm ext}$ we can also define an effective field theory operator $\tilde \Pi_\mathsf{R}$ that cyclically permutes the modes $\mathsf{R}$ between the $n$ copies of the state. Indeed, one way to define $\tilde \Pi_{n,\cal R}$ is simply as $\tilde V_{\rm ext}^{\dagger \otimes n} \Pi_n \tilde V_{\rm ext}^{\otimes n}$ where $\Pi_n$ permutes the $n$ external systems in the output of $\tilde V_{\rm ext}$. But it can also be defined directly in the effective field theory by the same methods described above. The $n$-th R\'{e}nyi ``swap entropy'' (in the language of \cite{Marolf:2020rpm}) can then be defined as
\begin{align}
    S_n(\mathsf{R}) = \frac{1}{1-n} \log \braket{\Psi|\tilde\Pi_{n,\mathsf{R}} |\Psi}~. 
\end{align}
Assuming the dominant gravitational saddle preserves the cyclic replica symmetry, one can use the Lewkowycz-Maldacena argument \cite{Lewkowycz:2013nqa, Faulkner:2013ana} to show that $S_n(\mathsf{R})$ is an analytic function of $n$ and that the von Neumann swap entropy
\begin{align} \label{eq:islandruleswap}
    S(\mathsf{R}) = \lim_{n \to 1} S_n(\mathsf{R}) = S_{\rm gen}(\mathsf{R} \cup I)~.
\end{align}
In the derivation of \eqref{eq:islandruleswap} it is important that one allows spacetime wormholes connecting different replicas when computing $S_n(\mathsf{R})$ \cite{Penington:2019kki,Almheiri:2019qdq}. For an ordinary quantum system (i.e., in the absence of gravity), it is clear that the von Neumann swap entropy is equal to the usual von Neumann entropy $S(\rho) = - \tr[\rho \log \rho]$. This is usually also assumed to be true in gravitational theories, although the presence of replica wormholes makes this conclusion less certain. In particular in theories of gravity that describe an ensemble average over quantum systems one has to be careful to distinguish the average entropy across the ensemble from the entropy of the ensemble-averaged state. The former is equal to the swap entropy \cite{Penington:2019kki, Bousso:2020kmy, Marolf:2020rpm}.

\section{Implications}
\label{sec-implications}

The gravitational path integral computation of Petz recovery implies that information in the island of $\mathsf{R}$ can be reconstructed from $\mathsf{R}$. But in a standard description of spacetime in terms of semiclassical gravity and effective field theory (SGEFT), spacelike observables commute. Thefore SGEFT cannot be valid in a region that includes $\mathsf{R}$ and any part of its island.

We have shown that the island lies outside of the stretched horizon for $\rho_R\ll \ell_p^{1/3} r_{\rm hor}^{2/3}$, and that it can extend as far as $\rho_I^{\rm max}\sim \sqrt{\ell_p r_{\rm hor}}$.\footnote{For simplicity, we specialize to $d=4$ spacetime dimensions in this section.} We conclude that SGEFT can be only be valid outside of $\rho_I^{\rm max}$. 

This is quite surprising. In any theory that allows information to be returned, SGEFT must break down in some way, since otherwise Hawking's original argument for information loss prevails~\cite{Hawking:1976ra}. For example, SGEFT must break down to allow a complementary description of the black hole interior in terms of exterior degrees of freedom, or to permit a structure like a firewall that prevents observers from entering the interior. However, it was widely believed that SGEFT could remain valid everywhere outside of the stretched horizon. 

Our results show that the breakdown of SGEFT must occur already at a much greater, potentially macroscopic distance from the horizon. We make this argument completely explicit in Sec.~\ref{sec-careful}. In Sec.~\ref{sec-verify} we show that the SGEFT-violating region can be probed by an asymptotic observer. In Sec.~\ref{sec-approaches}, we will discuss implications for various approaches to the black hole information paradox.

\subsection{Breakdown of Effective Field Theory Far Outside the Horizon}
\label{sec-careful}

Let $\sqrt{\ell_p r_{\rm hor}} \ll \rho_R \ll \ell_p^{1/3} r_{\rm hor}^{2/3}$, so that the island $I$ extends outside the stretched horizon, $\rho_I\gg \ell_p$. We suppose for contradiction that SGEFT is valid outside some radius $\rho_{\rm EFT}<\rho_I$.

The operator $\tilde V_{\rm ext}$ that extracts the modes $\mathsf{R}$ into an external quantum system acts within a causal diamond that remains outside of $\rho_R$. Hence, by assumption, $\tilde V_{\rm ext}$ is well described by effective field theory. In particular, $\tilde V_{\rm ext}$ commutes with the algebra associated with the island, and so after applying $\tilde V_{\rm ext}$ the external quantum system cannot encode any information that was supposed to be in the island. But the gravitational path integral predicts that Petz map reconstruction of island operators from the external system succeeds. This completes the contradiction.

Since $\rho_R$ depends parametrically on $G$, the above proof relies on our argument in Sec.~\ref{sec:mining} that backreaction from $\tilde V_{\rm ext}$ remains sufficiently small. Moreover, for a free theory $\tilde V_{\rm ext}$ has complexity $\exp[O(r_{\rm hor}^2/\rho_R^2)]$, which also depends on $G$.

In fact, to show that SGEFT must break down parametrically far outside the stretched horizon, it is sufficient consider $\rho_R = \epsilon r_{\rm hor}$ for arbitarily small but fixed $\epsilon \ll 1$ as $G \to 0$. As shown in Section \ref{sec:mining}, in this limit the existence of the asymptotic boundary operator $\tilde V_{\rm ext}$ that extracts all the modes $\mathsf{R}$ follows from a rigorous mathematical theorem and has complexity independent of $G$. In this case, we have $\rho_I \sim \ell_p^2 r_{\rm hor}^{-1} \epsilon^{-3} \ll \ell_p$ and so the island is inside the stretched horizon. But consider an ingoing $s$-wave photon with frequency $\omega = r_{\rm hor}/\epsilon$. The backreaction of this photon shifts the horizon by $\delta \rho \sim \ell_p^2 r_{\rm hor}^{-1} \epsilon^{-1} \ll \rho_I $. Suppose that at $t = 0$, the photon is at $\rho \sim \ell_p^2 r_{\rm hor}^{-1} \epsilon^{-5/2} \ll \rho_I$. 
It should therefore be encoded in the external quantum system after applying $\tilde V_{\rm ext}$ at $t = 0$.
\begin{figure}
    \centering
\includegraphics[width=0.5\textwidth]{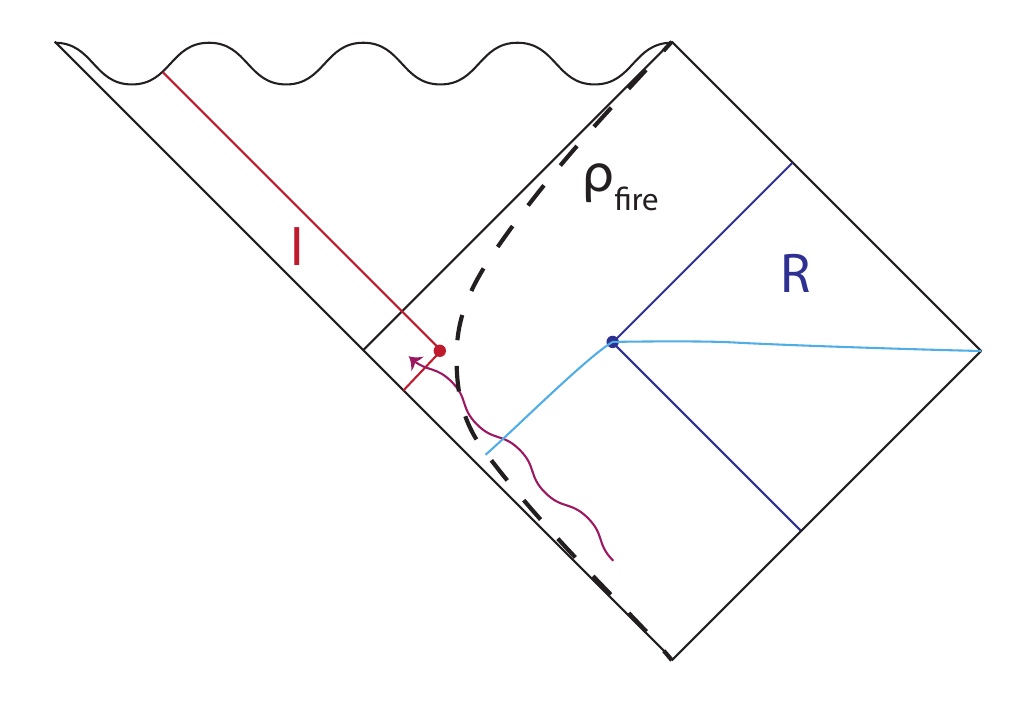}
    \caption{Even if we only consider regions $R$ with $\rho_R = \epsilon r_{\rm hor}$ for $\epsilon \ll 1$ independent of $G$, the location of the island in the semiclassical description is in conflict with the stretched-horizon firewall paradigm. We can find a partial Cauchy slice (light blue) in the exterior of the firewall that contains both an ingoing mode (red) that, in the semiclassical description, would enter the island (and that anyway is expected to be encoded in $\mathsf{R}$ based on the Hayden-Preskill decoding criterion) and the entirety of $\mathsf{R}$ as spacelike-separated degrees of freedom. As a result, the ingoing mode cannot be encoded in $\mathsf{R}$ without a cloning paradox.} 
    \label{fig:firewall}
\end{figure}
However at $t = - (\beta/2\pi) \log[r_{\rm hor} \epsilon^{2}/\ell_p]$, the photon will be at $\rho \sim \ell_p \epsilon^{-1/2}$. This can be made arbitrarily far outside the stretched horizon by dialing $\epsilon$. But it is also spacelike separated from $\rho_R = \epsilon r_{\rm hor}$ at $t = 0$. Thus a partial Cauchy slice contains both $\mathsf{R}$ and the photon while it is far outside the stretched horizon. This again leads to a contradiction between the assumed validity of SGEFT outside the stretched horizon, and the reconstructability of the island from $\mathsf{R}$ predicted by the gravitational path integral.

\subsection{Asymptotic Observers Can Verify the Breakdown}
\label{sec-verify}

Suppose a probe of the black hole is initially at rest just outside $\rho_{\rm EFT}\gg \ell_p$. The proper acceleration required for the probe to remain at rest is $\rho_{\rm EFT}^{-1}$. By slightly varying its acceleration, the probe can drop down to just inside $\rho_{\rm EFT}$ and then accelerate upwards again. To see this in detail, note that by the rocket equation, $m_0/m_1>e^{\tau/\rho}$, where $m_0$ and $m_1$ are the initial and final mass of the probe and $\tau$ is the duration of its flight. Gravitational backreaction will be small if $m_0\ll \rho_{\rm EFT}/G$; and the probe will be well-localized if $m_1\gg \rho_{\rm EFT}^{-1}$. Hence the probe can hover around $\rho_{\rm EFT}$ for about a local scrambling time, $\tau\ll \rho_{\rm EFT}\log(\rho_{\rm EFT}^2/G)$. This is an interesting limitation; but the time required to dive in and out of the EFT violating region is only of order $\rho_{\rm EFT}$.

A photon emitted by the probe midway through its flight, with local energy $\omega$, will reach infinity with energy $\omega \rho_{\rm EFT}/ r_{\rm hor}$. This is far higher energy than the background Hawking radiation if $\rho_{\rm EFT}^{-1} \ll \omega$. The local energy $\omega$ should be negligible compared to the probe mass at the time of emission, but this only leads to the weak constraint $\omega \ll M_p$. In order to pass through the potential barrier and reach infinity, the signal does need to have angular momentum $j^2 \lesssim \omega \rho_{\rm EFT}$. This is hard to achieve with high probability. But even a signal emitted with $O(1)$ angular precision will be transmitted with probability $p \sim \rho_{\rm EFT}^2 /r_{\rm hor}^2$, and so the signal can still be seen in semiclassical perturbation theory. Therefore the probe can communicate experimental results to a distant observer. In contrast, if $\rho_{\rm EFT} \lesssim \ell_p$, there is no way for asymptotic observers to ever detect the breakdown of SGEFT.

The probe described above can be created by an asymptotic boundary operator constructed using the timelike tube theorem (TTT). It therefore constitutes a method of probing $\rho < \rho_{\rm EFT}$ using only asymptotic boundary observables. This argument is of course somewhat abstract. In particular, it may not placate a reader who is skeptical (without clear justification) that the backreaction of local near-horizon operators constructed using the TTT is negligible. In Sec.~\ref{sec:mining} we gave a completely controlled argument establishing the breakdown of SGEFT at the $G$-independent distance $\rho_{\rm EFT}=\ell_p/\epsilon$ from the horizon; but placing the probe there by the TTT would again raise questions of backreaction.

Thus, we would ideally like to be able to directly probe $\rho < \rho_{\rm EFT}$ by a \emph{causal} process that can be carried out by distant observers inside the gravitating spacetime. This is indeed possible, so long as the distant observer has access to a NEC-saturating string with $O(1)$ tension in Planck units. Then a probe can be lowered slowly from infinity to $\rho_{\rm probe} < \rho_{\rm EFT}$. So long as $\rho_{\rm probe} \gg \ell_p$, it can then be pulled back out and read out.

Without NEC-saturating strings, probing the near-horizon region is harder but not impossible. One of the cleanest constructions for doing so that we have found is the following. Suppose we consider a black hole that is electrically charged with charge $Q$. So long as $Q \ll r_{\rm hor} \ell_p^{-1}$, the induced perturbation of the Schwarzschild metric will be small, and the results in this paper will still all apply. Moreover, if all charged particles in the theory are massive, their contribution to Hawking radiation will be exponentially suppressed for sufficiently large black holes, and so the black hole will remain charged after the Page time. Now suppose we construct a spaceship with mass $m$ and charge $q \gg m \ell_p$. This is easy to achieve if gravity is weak compared to electromagnetism: a positron, for example, has $q \approx 10^{-1}$ and $m \ell_p < 10^{-22}$. If the spaceship is at rest at $\rho_{\rm hover} \sim \epsilon \,r_{\rm hor}$, it will naturally want to accelerate outwards if $\epsilon \gtrsim G m M/ q Q$. Time-reversing this process, we can fire the spaceship towards the black hole at high speed so that it comes to a stop at $\rho_{\rm hover}$. 

Now the spaceship separates into a charged component and a neutral component. The charged component will then accelerate even more rapidly away from the black hole. Because the potential energy accumulated during the fall is stored in the electric field, only the mass of the remaining probe is left behind, which is redshifted by a factor $\epsilon$.
This is a key advantage over using a rocket to end up hovering in the same location: the rocket exhaust would fall into the black hole, potentially hiding the island behind the horizon, whereas the charged component escapes to infinity. 
The total energy $E$ of the emitted bremsstrahlung scales as $E \sim q^2 a^2 \tau \sim q^2/r_{\rm hor}$, where $a$ is the proper acceleration of the charge $q$ and $\tau$ is the time it lasts. If this radiation falls into the black hole, the resulting backreaction on the horizon will be small: $\delta \rho_{\rm hor} \sim G r_{\rm hor}^{-1} \ll \ell_p$.

We are left with a neutral spaceship at rest at $\rho_{\rm hover} \ll r_{\rm hor}$. This will free-fall into the black hole. The proper time between the spaceship reaching $\rho_I^{\rm max} \sim r_{\rm hor}^{1/2} \ell_p^{1/2}$ and crossing the horizon will be $T_{\rm island} \sim \ell_p r_{\rm hor}/ \rho_{\rm hover} \gg \ell_p$. This is plenty of time to send a message back out to infinity. The backreaction on the horizon from the spaceship falling in will be $\delta \rho_{\rm hor} \sim \ell_p m^{1/2} \rho_{\rm hover}^{1/2}$, which is much smaller than $\rho_I^{\rm max}$ for $m \sim M_p$. A photon of frequency $\omega$ in the spaceship's rest frame will reach infinity with energy $E \sim \omega \ell_p / \rho_{\rm hover}$ which is parametrically higher than the background Hawking radiation for $\omega \gg \ell_p^{-1} \rho_{\rm hover}/ r_{\rm hor}$.

\subsection{Implications for the Information Paradox}
\label{sec-approaches}

We now discuss the implications of the breakdown of SGEFT far from horizons for a number of approaches to the black hole information paradox.

\paragraph{Complexity-protected complementarity} Approaches such as ER=EPR and nonisometric maps concede a breakdown of SGEFT when certain states arise in $\mathsf{R}$ that are sufficiently computationally complex. In their absence, infalling observers are described by SGEFT to a good approximation; but in principle, it is possible to manipulate the black hole interior from $\mathsf{R}$. Before our result, these highly nonlocal effects could only be verified by observers who entered the black hole. Now it is clear that they can also be explored, at least in principle, by an asymptotic observer.

It is worth being precise here about the sense in which an asymptotic observer is able to manipulate $\mathsf{R}$ to change the state of the island and then verify those effects. It seems highly implausible that an observer within the gravitating spacetime could actually implement the exponentially complex manipulations required in real time. But it is also implausible that an observer within the gravitating spacetime could actually implement the AMPS experiment. The most compelling response to this objection in the context of AMPS is that a boundary observer in AdS/CFT (or whatever the equivalent of that observer is in flat space) could extract the Hawking modes out of the spacetime, manipulate them as desired and then put them back, before they (or a different bulk observer) jumps into the black hole. One might object here that of course holography implies that a boundary observer can change the interior of the black hole by acting on the CFT. But, crucially, the extraction of the radiation does not touch the black hole interior: it just extracts some modes that were already well outside the horizon.

The discussion here is almost exactly analogous, except that there is no need for an observer to ever jump into the black hole. The asymptotic observer first extracts the modes $\mathsf{R}$ using $\tilde V_{\rm ext}$, which in SGEFT would not affect the state of the island. (This temporarily hides the island behind a horizon.) They then implement a Petz map reconstruction of an island operator on the extracted modes. From the point of view of SGEFT, the resulting state is unchanged except for the creation of a low-energy excitation in the island. Finally they apply $\tilde V_{\rm ext}^\dagger$, putting the modes back in the spacetime and thereby undoing the backreaction and revealing the island again in its new state. This new state can then be verified by the methods described in Section \ref{sec-verify} above.

\paragraph{Firewalls} AMPS proposed that infalling observers burn up at the stretched horizon~\cite{Almheiri:2012rt}, that effective field theory remains valid in its exterior, and that the Hawking radiation returns the information that went into the black hole. This would alter physics sharply at the horizon, but it promised to obviate the need for complimentarity of any sort. 

However, if there is indeed a firewall at the stretched horizon, then our result implies that SGEFT breaks down outside of it. Thus, a firewall at the stretched horizon would fail to eliminate the need for complementarity.

A possible alternative is to posit that the firewall location coincides with the boundary of the largest island, at a distance of order $\rho_I^{\rm max} \sim \ell_p^{1/2} r_{\rm hor}^{1/2}$ from the horizon. This would imply that SGEFT fails to compute even simple boundary observables in AdS/CFT. More generally, it would imply that effective field theory fails to predict the outcome of simple experiments conducted by distant observers, such as lowering a probe sufficiently close to (but much farther than $\ell_p$ from) the horizon.

This is a radical modification of physics even compared to a firewall at the stretched horizon. Moreover, a key lesson of islands is that SGEFT is able to predict even nonperturbatively small contributions to boundary observables (specifically to replica trick permutation operators) that are required for boundary unitary. It is hard to understand why SGEFT would succeed at that while failing to correctly predict simple observables at finite, or even leading, order in perturbation theory. In particular, one possibility that has been explored in recent years \cite{Bousso:2019ykv,Marolf:2020rpm,Marolf:2020xie, Bousso:2020kmy} is that semiclassical gravity (without firewalls) is holographically dual to the average over an ensemble of unitary boundary theories, each of which individually has firewalls. This cannot work if the firewalls need to be far outside the stretched horizon: if a signal seen at infinity in SGEFT is absent in each of the theories with firewalls, it cannot appear after averaging over those theories.

One might be tempted to restore the validity of SGEFT for computing boundary observables, by positing that the dynamics of the far-out firewall are fine-tuned to reproduce the predictions from SGEFT. For example, an observer lowered into the firewall on a string would have to be re-emitted later with memories of seeing empty space filled with Hawking radiation. (Roughly, this is analogous to positing that the universe was created yesterday with initial conditions matching our memories and present observations.) However, for this to be true the firewall dynamics would need to have a sophisticated dual holographic description of at least the region $\ell_p \ll\rho < \rho_{\rm EFT}$. By the arguments above, that dual description would have to involve black hole complementarity, so we are led back to possibilities already considered above.

\paragraph{Other approaches} The AMPS mining argument shows that all high angular momentum Rindler modes in the black hole atmosphere carry information and so cannot be in the vacuum. This implies the excitation of all corresponding horizon-spanning (non-Rindler) modes in the freely falling frame, and hence the termination of infalling observers, at a distance from the horizon corresponding to the highest possible UV cutoff. 

The most conservative option available to AMPS was then to propose that a firewall is localized at the stretched horizon. This had the advantage of not modifying the coarse-grained outside description of the black hole, compared to approaches such as non-violent nonlocality~\cite{Giddings:2012gc} and (some versions of) fuzzballs~\cite{Bena:2022rna}. In light of islands far outside the horizon, this option is no longer available. 

But of course, our result rules out any description of the black hole in which SGEFT is asserted to hold outside of the stretched horizon (e.g.~\cite{Guo:2017jmi}), regardless of whether or not this assertion was motivated by the AMPS mining argument.

\subsection*{Acknowledgements}
This work was supported in part by the Berkeley Center for Theoretical Physics; by the Department of Energy, Office of Science, Office of High Energy Physics under QuantISED Award DE-SC0019380 and under contract DE-AC02-05CH11231. RB was supported by the National Science Foundation under Award Number 2112880. GP was supported by the Department of Energy through an Early Career Award DE-FOA-0002563 and by AFOSR award FA9550-22-1-0098.

\bibliographystyle{JHEP}
\bibliography{covariant}

\end{document}